\documentclass[longauth]{aa} 
\usepackage{graphicx}
\usepackage{txfonts}
\begin{document}

   \title{The nature of LINER galaxies:}
   \subtitle{Ubiquitous hot old stars and rare accreting black holes}

   \author{R. Singh			\inst{1}
		\thanks{Member of IMPRS for Astronomy \& Cosmic Physics at the University of Heidelberg} \and
		G. van de Ven		\inst{1}	\and
		K. Jahnke			\inst{1}	\and
		M. Lyubenova		\inst{1}	\and
		J. Falc\'on-Barroso	\inst{2,3}	\and
		J. Alves			\inst{4}	\and
		R. Cid Fernandes	\inst{5}	\and
		L. Galbany		\inst{6}	\and
		R. Garc\'ia-Benito 	\inst{7}	\and
		B. Husemann		\inst{8}	\and
		R. C. Kennicutt		\inst{12}	\and
		R. A. Marino		\inst{13}	\and
		I. M\'arquez		\inst{7}	\and
		J. Masegosa		\inst{7}	\and
		D. Mast			\inst{7,9}	\and
		A. Pasquali		\inst{10}	\and
		S. F. S\'anchez		\inst{7,9}	\and
		J. Walcher		\inst{8}	\and
		V. Wild			\inst{11}	\and
		L. Wisotzki		\inst{8}	\and	
		B. Ziegler			\inst{4}	\and
        		the CALIFA collaboration}

   \institute{Max-Planck-Institut f\"ur Astronomie (MPIA), K\"onigstuhl 17, 69117 Heidelberg, Germany  	
             	\and Instituto de Astrof\'isica de Canarias (IAC), E-38205 La Laguna, Tenerife, Spain 		
         	\and Depto. Astrof\'isica, Universidad de La Laguna (ULL), 38206 La Laguna, Tenerife, Spain	
		\and University of Vienna, T\"urkenschanzstrasse 17, 1180 Vienna						
		\and Departamento de F\'isica, Universidade Federal de Santa Catarina, PO Box 476, 88040-900 Florian\'opolis, SC, Brazil		 
		\and CENTRA - Centro Multidisciplinar de Astrof\'isica, Instituto Superior T\'ecnico, Av. Rovisco Pais 1, 1049-001 Lisbon, Portugal	
		\and Instituto de Astrof\'isica de Andaluc\'ia (CSIC), Glorieta de la Astronom\'ia s/n, E18008 Granada, Spain	
		\and Leibniz-Institut f\"ur Astrophysik Potsdam (AIP), An der Sternwarte 16, D-14482 Potsdam, Germany		
		\and Centro Astron\'omico Hispano Alem\'an, Calar Alto, (CSIC-MPG), C/Jes\'us Durb\'an Rem\'on 2-2, E-04004 Almer\'ia, Spain	
		\and Astronomisches Rechen Institut, Zentrum f\"ur Astronomie der Universit\"at Heidelberg, M\"onchhofstrasse 12-14 , D-69120 Heidelberg, Germany	
		\and School of Physics and Astronomy, University of St Andrews, North Haugh, St Andrews, KY16 9SS, U.K. (SUPA)	
		\and Institute of Astronomy, University of Cambridge, Madingley Road, Cambridge CB3 0HA UK	
		\and CEI Campus Moncloa, UCM-UPM, Departamento de Astrof\'{i}sica y CC. de la Atm\'{o}sfera, Facultad de CC. F\'{i}sicas, Universidad Complutense de Madrid, Avda.\,Complutense s/n, 28040 Madrid, Spain 
		}

\titlerunning{The nature of LINER galaxies}
\authorrunning{Singh et al.}

   \date{Received 12/06/2013; accepted 08/08/2013}

  \abstract  
  {
  Galaxies, which often contain ionised gas, sometimes also exhibit a so-called low-ionisation nuclear emission line region (LINER).
  For 30 years, this was attributed to a central mass-accreting supermassive black hole (more commonly known as active galactic nucleus or AGN) of low luminosity, 
  making LINER galaxies the largest AGN-sub-population, which dominate in numbers over higher
  luminosity Seyfert galaxies and quasars.
  This, however, poses a serious problem. While the inferred energy
  balance is plausible, many LINERs clearly do not contain
  any other independent signatures of an AGN.}
    {Using integral field spectroscopic data from the CALIFA survey, we compare the observed radial surface brightness profiles with what is expected from illumination by an AGN.}
  {Essential for this analysis is a proper extraction of emission lines, especially weak lines, such as Balmer H$\beta$ lines which are superposed on an absorption trough. To accomplish this, we use the GANDALF code, which simultaneously fits the underlying stellar continuum and emission lines.}
  {For 48 galaxies with LINER-like emission we show, that the
  radial emission-line surface brightness profiles are inconsistent with
  ionisation by a central point-source and hence cannot be due to an AGN
  alone.}
  {The most probable explanation for the excess LINER-like
  emission is ionisation by evolved stars during the short but very hot and
  energetic phase known as post-AGB.
  This leads us to an entirely new interpretation.
  Post-AGB stars are ubiquitous and their ionising effect should be potentially
  observable in every galaxy with the gas present and with stars older than $\sim$1
  Gyr unless a stronger radiation field from young hot stars or an AGN outshines
  them. This means that  galaxies with LINER-like emission are not a
  class defined by a property but rather by the absence of a property.
  It also explains why LINER emission is observed mostly in
  massive galaxies with old stars and little star formation.
  }
  {}


   \maketitle
%

\section{Introduction}
\begin{figure*}[t]
	\includegraphics[width=\hsize]{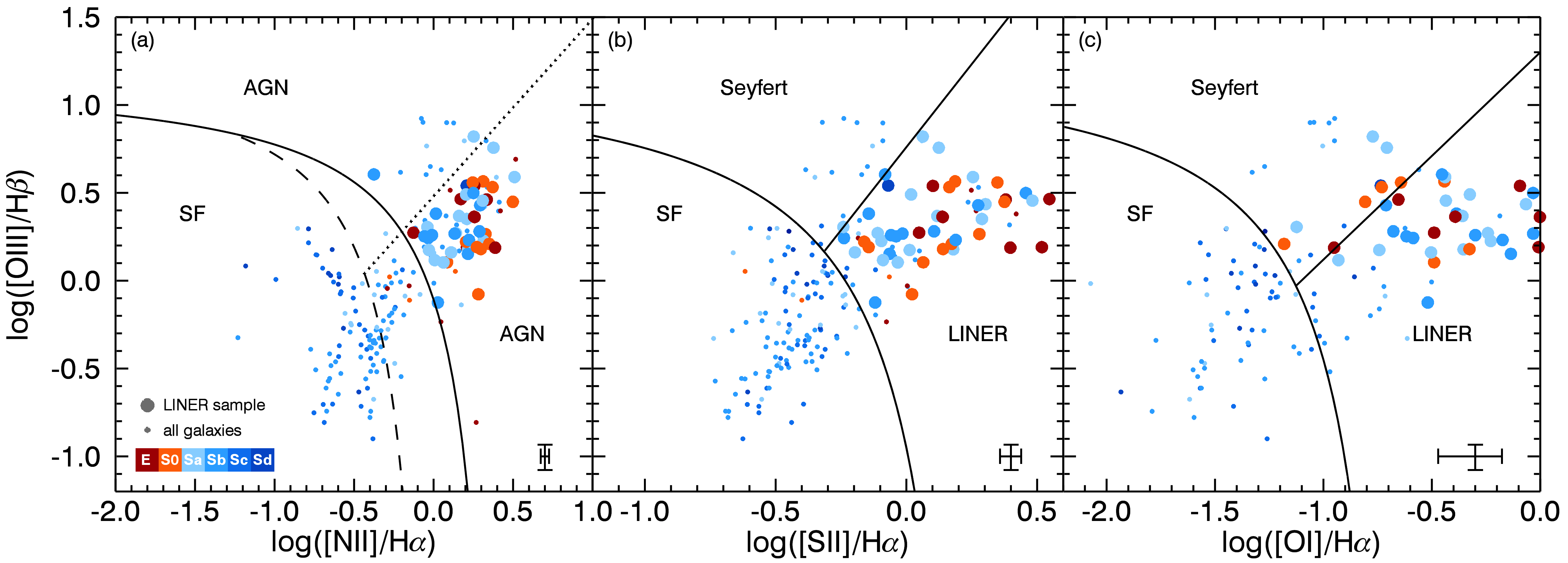}
	\caption{\textbf{Selection of LINER galaxies.} 
	Data points are coloured by Hubble type and show central emission line ratios of 257 CALIFA galaxies on the BPT diagram. Those classified as AGN in (a) and LINER in (b) have larger symbol sizes and are selected for further analysis. Small symbol sizes in the LINER region may appear due to either an inconsistent BPT classification, this typically affects points near the demarcation lines, which are either LINERs in (b) but not AGN in (a), vice versa, or objects that have had the central measurement detect only one of either [NII] or [SII] emission lines. Panel (c) is shown for illustration but not used in the classification due to the larger relative error indicated in the lower right corner of each panel. The solid curves are the theoretically modelled "extreme starburst line" \citep{kewley2001}. The dashed curve \citep{kauffmann2003} and the dotted line \citep{cidfernandes2010} in (a) and the solid lines \citep{kewley2006} in (b) and (c) are tracing a minimum in the central emission-line-ratio-distribution of SDSS galaxies.}
	\label{fig:fig1}
\end{figure*}

When LINERs were first identified as a class of galaxies in the early 1980s,
it was clear that the necessary radiation field had to be different or have a
different impact than for all previously known AGNs.
For both the rare high-power quasars observable across the whole visible
Universe and the more common Seyfert galaxies, well-understood
models of accretion disks could be computed \citep{shakura1973, narayan1994}. Various explanations for the
LINERs were put forward, ranging from shock-ionisation \citep{heckman1980} via
young hot stars \citep{terlevich1985} to the favoured ionisation by
low-luminosity AGNs \citep{ferland1983,halpern1983}. The latter explanation has
strong implications, because LINERs make up most objects in the AGN class.
While in the following decades the explanation that LINERs are powered by
low-luminosity AGNs became generally accepted,
doubts were again re-fuelled very recently.
Inconsistencies were found between the AGN-ionisation hypothesis,
and either predicted emission line strengths \citep{cidfernandes2011} or the
spatial distribution of LINER-like ionised regions in the
galaxies \citep{sarzi2010,yan2012}, but neither were conclusive, because
  they either lacked full spatial or spectral information.

In this work, we base our analysis on a new dataset that combines a complete spectral and spatial view on LINER galaxies for the first time.
We test the picture of ionisation of the gas in these galaxies by an AGN as our null-hypothesis,
which geometrically defines illumination by a single central point source.
This geometry predicts a radiation field declining in radius as $\propto
1/r^2$.
With interstellar gas density in galaxies normally distributed in a thin disk
with an exponential fall off in radius \citep{bigiel2012}, the density of
ionised gas and hence surface brightness of emission line flux should
also fall off similar to $1/r^2$ or faster, \emph{if}
LINER-like emission across the galaxy were caused by a central
AGN point-source.
If the surface brightness of spatial regions with LINER-like emission falls
off less steeply than $\propto 1/r^2$, then it is not reconcilable with
illumination by an AGN.
 
For this test, both the full spatial resolution of the galaxies 
as well as the ability to spectrally identify LINER-like emission
by means of diagnostic emission line ratios \citep{heckman1980} is required.
Whereas either no full spatial component was
available \citep{cidfernandes2011,yan2012} or spectral
coverage for the LINER diagnostic was
limited before \citep{sarzi2010}, the CALIFA
survey \citep{sanchez2012,husemann2012} provides the first
dataset of this kind for a substantial number of LINER galaxies \citep[see also][]{kehrig2012}.

In Section \ref{sec:CALIFA}, basic information about the CALIFA survey is provided.
In Section \ref{sec:analysis}, we describe our sample selection and the steps of our analysis.
In Section \ref{sec:tests}, we discuss different effects, which might influence our results and show that they are robust against them.
We conclude in Section \ref{sec:discussion} and discuss the implications of our results.

\section{The CALIFA survey}
\label{sec:CALIFA}
The Calar Alto Legacy Integral Field Area \citep[CALIFA; ][]{sanchez2012} survey is the first and ongoing IFS survey of a diameter-selected ($45\arcsec < D_{25} < 80\arcsec$) sample of up to 600 galaxies in the local universe ($0.005 < z < 0.03$) of \emph{all Hubble types}. 
The data are being obtained with the integral-field spectrograph PMAS/PPak mounted on the 3.5 m telescope at the Calar Alto observatory. Its field-of-view of $65\arcsec \times 72\arcsec$ covers the full optical extent of the selected galaxies.
More than 96\% of the CALIFA galaxies are covered to at least two effective radii. The coverage distribution peaks around 4 $R_e$ and reaches up to 7 $R_e$ for highly inclined galaxies.

This survey comprises two different gratings for each galaxy: one at a lower spectral resolution (V500) of 6.0 \AA \, FWHM and one at a higher resolution (V1200) of 2.3 \AA \, FWHM. Since our specific analysis requires almost the whole optical wavelength range, only the V500 data are used, which cover a nominal wavelength range of 3745--7500 \AA \, at a median spatial resolution of 3\farcs7.

The medium-resolution V1200 grating yields high-quality maps of stellar and ionised gas kinematics. The combination with the low-resolution V500 grating allows for mapping of stellar ages, metallicities, full star-formation histories, ionised-gas emission line fluxes, and chemical abundances \citep[e.g. ][]{perez2013, sanchez2013,FalconBarroso_inprep}.

The minimal S/N of the spatially-binned V500 spectra used in this analysis was set to 10. Since most spaxels have much higher S/N values, binning was only necessary in the galaxies peripheral areas. 

\section{Spatially resolved LINER-like emission}
\label{sec:analysis}
Among the first 257 galaxies observed within the CALIFA survey, we found 48
LINER galaxies based on their measured central emission line ratios, covering a 3" diameter aperture.
These galaxies cover almost all morphological types based on the averages of five independent visual classifications of SDSS \textit{r} and \textit{i} band images.
Based on the ionisation strength and hence the underlying ionisation source, the so-called BPT diagram \citep{baldwin1981} is an empirically derived diagnostic tool to distinguish between star formation (dominated by Balmer H$\alpha$ and H$\beta$ lines), Seyfert galaxies (with high ionisation potential) and LINER galaxies (with low ionisation lines).

As shown by this diagnostic diagram in Fig.~\ref{fig:fig1}, the flux in the
lower-ionisation lines [NII]$\lambda$6583, [SII]$\lambda \lambda$6716,31
and [OI]$\lambda$6300 when compared to H$\alpha$$\lambda$6563 is too high
for ionisation by young massive stars, and at the same time, the flux in
[OIII]$\lambda$5007 when compared to H$\beta$$\lambda$4861 is too low for
ionisation around a typical Seyfert-like medium-luminosity AGN.
The emission lines that are combined in the ratios are close in wavelength;
therefore, attenuation by dust cancels out, leaving the galaxies to be
classified as LINERs.

\begin{figure*}[t]
   	\includegraphics[width=\hsize]{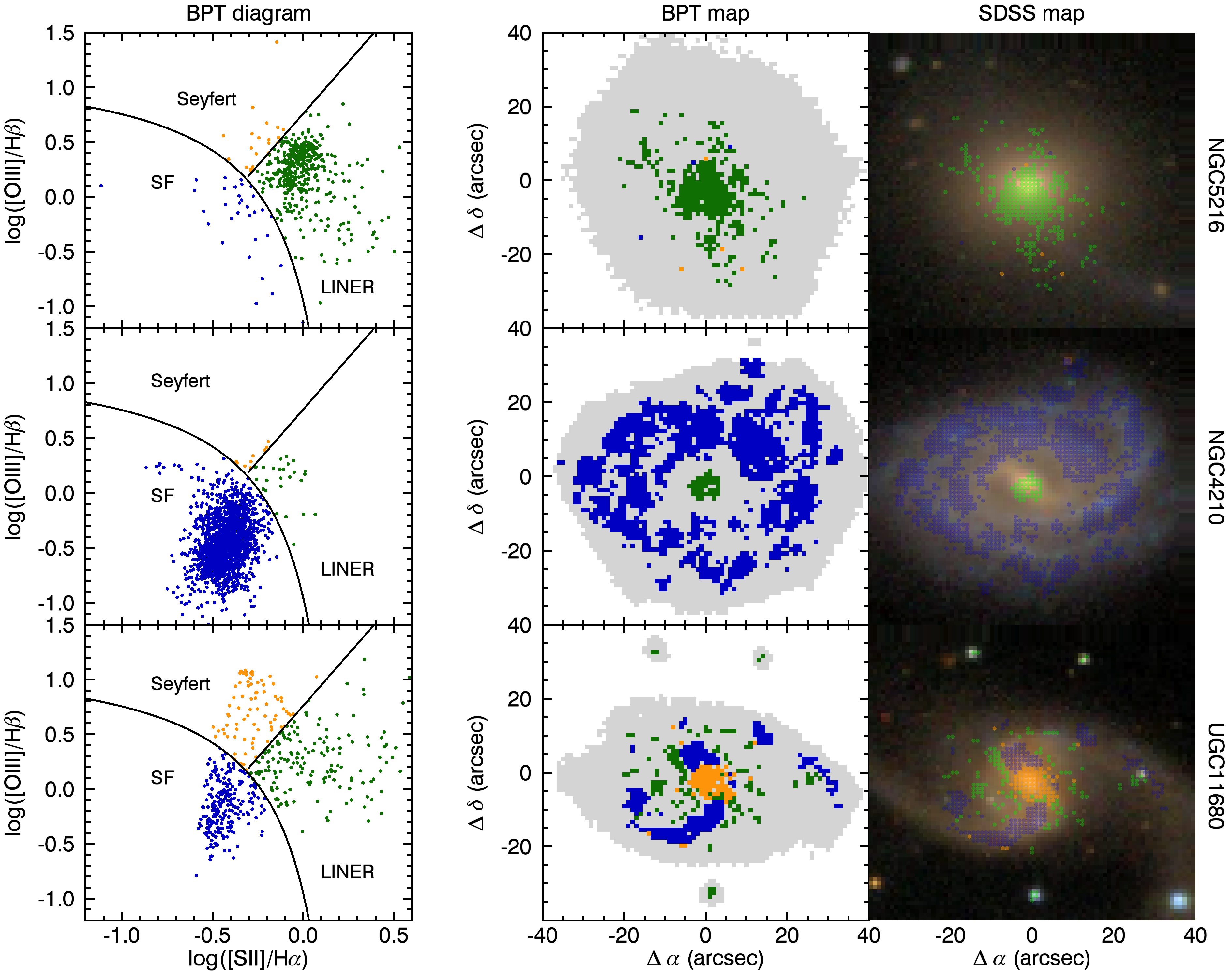}
   	\caption{
	  \textbf{Emission-line-ratio classification of spatial regions.}
          The distribution of measured emission line ratios in a BPT-diagram (left column) and
          spatially across the galaxy (middle column, usable data regions in grey) 
          differ between example galaxies dominated by LINER-like emission
          (NGC\,5216, top row), a galaxy dominated by star-formation (NGC\,4210, middle row)
          and an AGN in a spiral host (UGC\,11680, bottom row).
          The latter one is shown as an example but not part of our sample.
	Regions are colour-coded according to their position in the BPT diagram: green for LINER-like, orange for Seyfert-like, and blue for star-formation-like emission line ratios.
	Overlay of these regions onto the colour-composite image from the SDSS (right column) reveals how LINER-like emission is spatially extended, except when dominated by Seyfert-like emission in the centre or by star-formation-like emission in spiral arms or the disk.
        }
	\label{fig:fig2}
\end{figure*}

Subsequently, we measured emission line ratios in different regions across the galaxy
for each of these LINER galaxies.
Even the weak emission lines can be robustly recovered as we are simultaneously
fitting the stellar continuum and emission lines, while requiring a minimum signal-to-noise
of 10 per pixel; in the outer parts we combine spectra from neighbouring regions to
reach this minimum signal-to-noise.
As illustrated for three example galaxies in Fig.~\ref{fig:fig2},
we can then place all regions with reliable measured line ratios on the BPT
diagrams and classify each of them.
Next, we can colour-code all regions across the galaxy according
to this classification to obtain BPT maps. This map reveals which parts of
the galaxy are dominated by star-formation-like emission (typically in
the outer parts and/or in spiral arms), Seyfert-like emission (restricted to the
centre), or LINER-like emission (typically extended well beyond the nucleus).

\begin{figure*}[t]
 \includegraphics[width= \hsize]{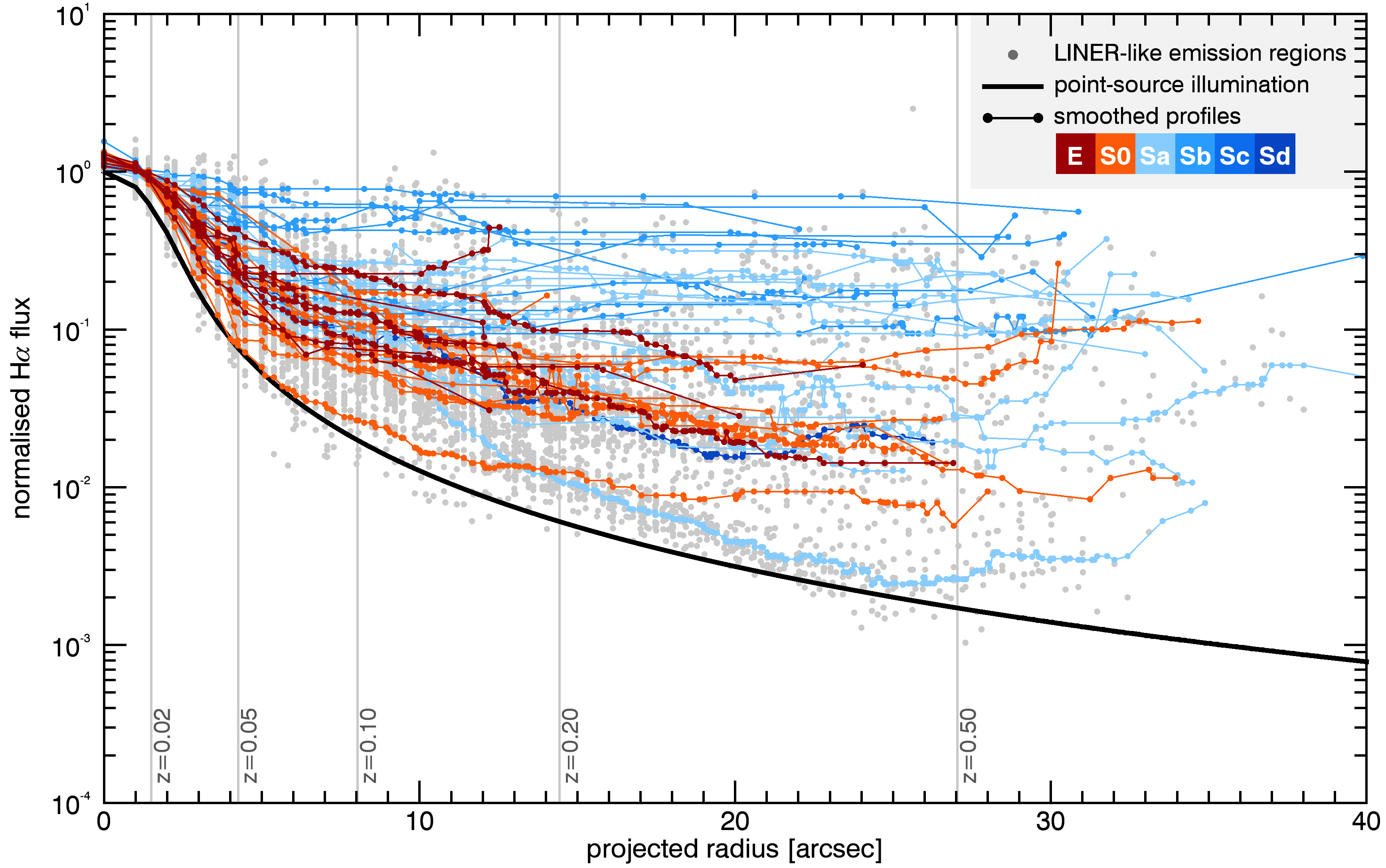}
 \caption{\textbf{Radial profiles of LINER-ionised H$\alpha$ flux.} 
The coloured curves are normalised and smoothed radial surface brightness
profiles of the H$\alpha$ emission line flux from our 48 LINER
galaxies and compared to a PSF convolved point-source illumination $1/r^2$-profile in black. All 
profiles are normalised with respect to the central flux inside a 1\farcs5 
radius aperture. Different colours represent different morphological types
ranging from round elliptical galaxies (dark red) to disk-dominated spiral
galaxies (dark blue), with lighter colours in-between. 
Beyond the inner few arcsec for LINER galaxies of all types,
there is a significant excess, with up to two orders of magnitude, above the
prediction from a point-source ionisation. 
The vertical grey lines illustrate the radial extent covered by a 3"-diameter SDSS aperture when the CALIFA galaxies with $ \bar{z} =0.017 \pm 0.006$ would be placed at the indicated higher redshifts.
}
 \label{fig:fig3}
\end{figure*}

Finally, we select only those regions with LINER-like emission and plot the
measured H$\alpha$ surface brightness -- or that of any other emission
line -- versus the distance of each region from the galaxy centre. After
normalising the central H$\alpha$ flux to unity for all galaxies, we arrive at the
(smoothed) coloured radial profiles in Fig.~\ref{fig:fig3}. The
expected profile from central point-source illumination is plotted in black.
There is a strong gap between the latter predicted point-source-illumination
profile and the actual observed profiles,which increase to $\gtrsim 1$\,dex
toward the $\sim 30$" radial extent of our data.
In some regions, part of the emission can be the result of a superposition of
different ionising flux sources. In Section \ref{sec:tests}, we
describe tests, which assure that the discrepancy between data and the null
hypothesis model is not due to this contribution. These tests show that our
results are not affected by 
a potential contribution to the line flux triggered by young stars.
The vertical grey lines in Fig. 3 illustrate the radial extent covered by a 3"-diameter SDSS aperture, if the CALIFA galaxies with an average redshift of 0.017 $\pm$ 0.006 would be placed at the indicated higher redshifts. This illustrates that SDSS emission line classifications can be non-unique and dependant on redshift, meaning that for example NGC\,4210 from Fig. 2 could be classified as a either star-forming or LINER galaxy, depending on its distance and hence, apparent size.

In addition, we tested against projection effects for the disk-dominated
galaxies in our sample, which affects both radius coordinates and flux
densities. After these tests the discrepancy between observed radial line
surface brightness profiles and the null hypothesis remains for both
early- and late-type galaxies, hence rejecting the model that a central AGN
is causing the spatially extended LINER-like emission.


\section{Verifying the robustness of our analysis}
\label{sec:tests}

Below, we discuss various effects that could influence our results and verify that our findings are robust against them.

\subsection{Robustness of weak emission line extraction}

A central technical part of this work is the extraction of weak emission
lines in the presence of a stellar continuum. Since the equivalent width of
the lines is often low, specifically with the Balmer H$\beta$ line
being superposed on an absorption trough, an
unbiased extraction is the core of the present analysis.

We have employed three different procedures for line extraction. All
avoid to model the line flux after a previous subtraction of
the stellar continuum. Instead, continuum and line emission are modelled
simultaneously, which provides the least-biased line
measurement \citep{sarzi2006}. To assess whether systematics in the
line measurement are present and whether the uncertainties on line fluxes -- and
hence line ratios -- are properly derived, we compared the extracted line fluxes
and errors from the following complementary approaches: 

(a) We use the \textbf{g}as \textbf{and} \textbf{a}bsorption-\textbf{l}ine \textbf{f}itting procedure GANDALF \citep{sarzi2006} with the MILES \citep{sanchez2006, falconbarroso2011} library of stellar templates.
In this case, the best fit to a spectrum is the superposition of an optimal combination of the stellar
templates with additional Gaussians representing the emission
lines. Unfortunately, the GANDALF routine only computes errors for those
emission lines that are unrestricted in their kinematical properties. Emission
lines that are being tied to another line in either velocity or velocity
dispersion are better recovered \citep{sarzi2006} but do not come with errors
for the measured fluxes.

(b) Same as (a) but we used the MILES library of single stellar populations (SSP) instead of the MILES stellar template library.

(c) To acquire flux errors for all emission lines of interest, we employed a
Monte-Carlo simulation, perturbing the input spectrum one hundred times. This
amount of different realisations is sufficient to create a Gaussian
distribution in extracted fluxes from which we take the mean and
standard deviation as measured flux and error value. To make this process
computationally feasible, we use the best-fit composite stellar spectrum from a previous emission-line masked stellar continuum fit with the procedure PPXF \citep{cappellari2004} instead of the full template library.

\begin{figure}[t]
\includegraphics[width= \columnwidth]{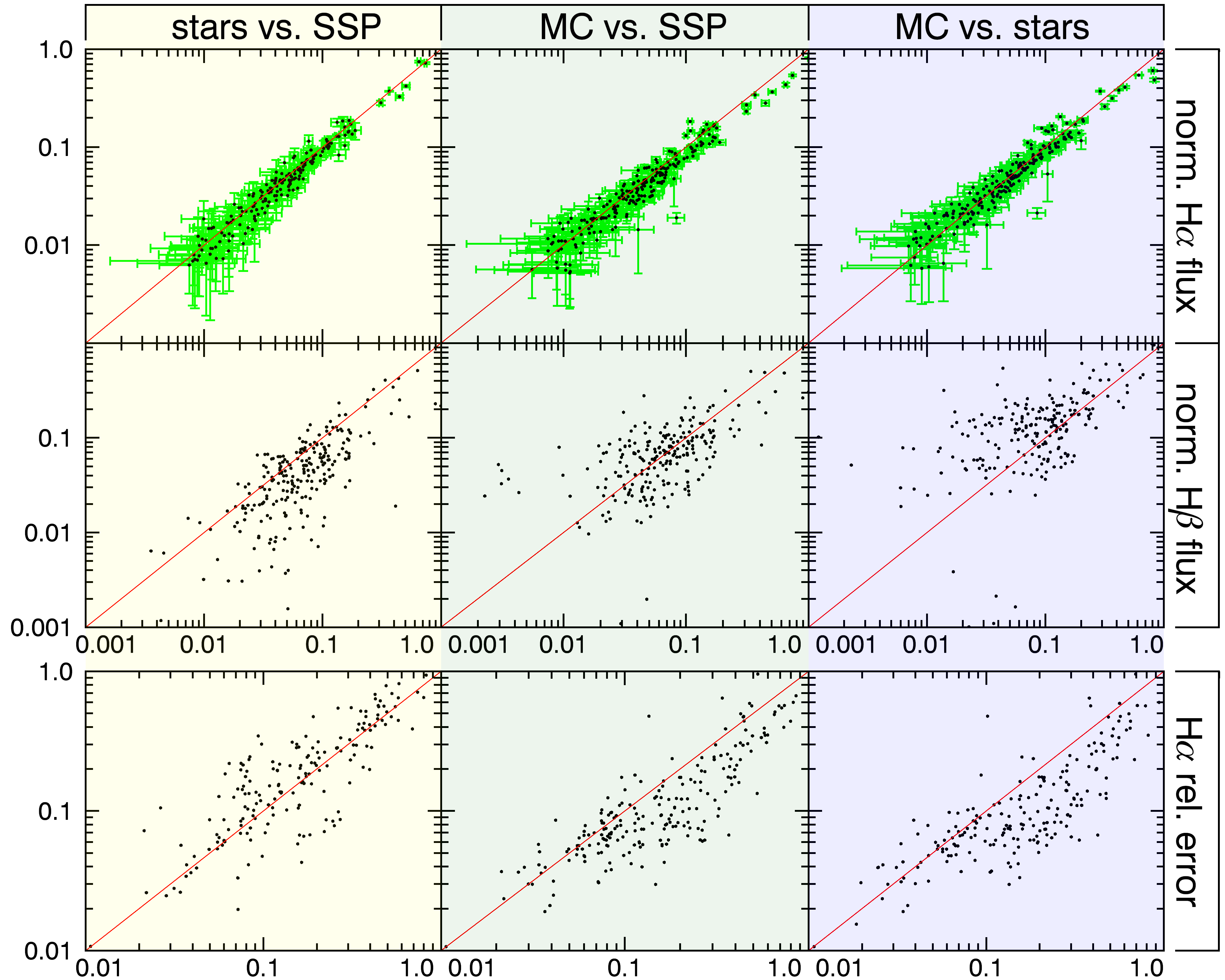}
\caption{Comparison of extracted H$\alpha$ fluxes, H$\beta$ fluxes and estimated errors for the
  H$\alpha$ line of NGC\,5614, as extracted by Gandalf using the MILES stellar template library, the SSP
  library and a Monte Carlo variation of noise in the spectra. There is a very good match
  in the line fluxes between the three methods for the H$\alpha$ line (row 1), while the errors estimated
  by Gandalf are larger than the statistical variance from the Monte Carlo approach (row 3).
  The weak H$\beta$ line is generally better recovered using either the Monte-Carlo or GANDALF/SSP method (row 2).
  }
	\label{fig:faintextraction}
\end{figure}

As an example in Fig.~\ref{fig:faintextraction}, the resulting fluxes and
errors are compared for the H$\alpha$ and H$\beta$ line of NGC\,5614. 
There is no systematic difference in flux extraction between the three methods also at the faint
end with a scatter within each method's uncertainty.
The $\chi^2$-based error from GANDALF, most likely due to unaccounted small pixel-to-pixel systematics and correlations, are larger than the robust Monte Carlo measurements. We verified that even if the Monte Carlo errors happened to be underestimated our results are unchanged, and we conclude that the line flux properties are accurately measured, down to the faintest end.

\subsection{Point source radiation to radial flux profile}

The radiation field from a central point source like an AGN declines with the
inverse of the radial distance $\propto 1/r^2$. In the case that the photo-ionised gas
is optically thin and distributed in an infinitesimally thin disk with a
constant filling factor and constant density, the resulting observed
emission-line flux also falls off inversely squared with (projected)
radius $\propto 1/R^2$.

The fall-off is even faster when the gas is not optically thin and part
of the ionisation gets absorbed by intervening gas
(clouds). Similarly, a radially decreasing filling factor results in a
faster decline, whereas the opposite of an increase would require very special
conditions.
With perhaps the exception of strongly interacting galaxies, the gas density
in galaxies is normally radially decreasing \citep{bigiel2012}, so that the flux is
also expected to drop at an even faster gradient than inverse square in this case.
Only in the case that the thin-disk assumption is strongly invalidated do we expect
the opposite effect of a decline shallower than $\propto 1/R^2$ -- in the
extreme case of optically thin gas with a constant filling factor and constant
density in a spherical distribution, the line-of-sight integral results in an
observed emission-line flux that will fall off inversely linear with projected
radius.

However, the resulting kinematics in all types of galaxies with (ionised) gas present
shows clear disk-like rotation, apart from disturbances due to
non-axisymmetric structures (bars and spiral arms) and tidal interactions
(e.g., Garcia-Lorenzo et al., in prep.).
The resulting angular momentum implies that the gas always settles in a
disk, which, however, can have a substantial thickness. After line-of-sight
integration, the latter thickness still results in a slightly slower
fall-off than the inverse square, but the radially declining gas density
typically compensates for this.

\begin{figure}
\begin{center}
	\includegraphics[width=\columnwidth]{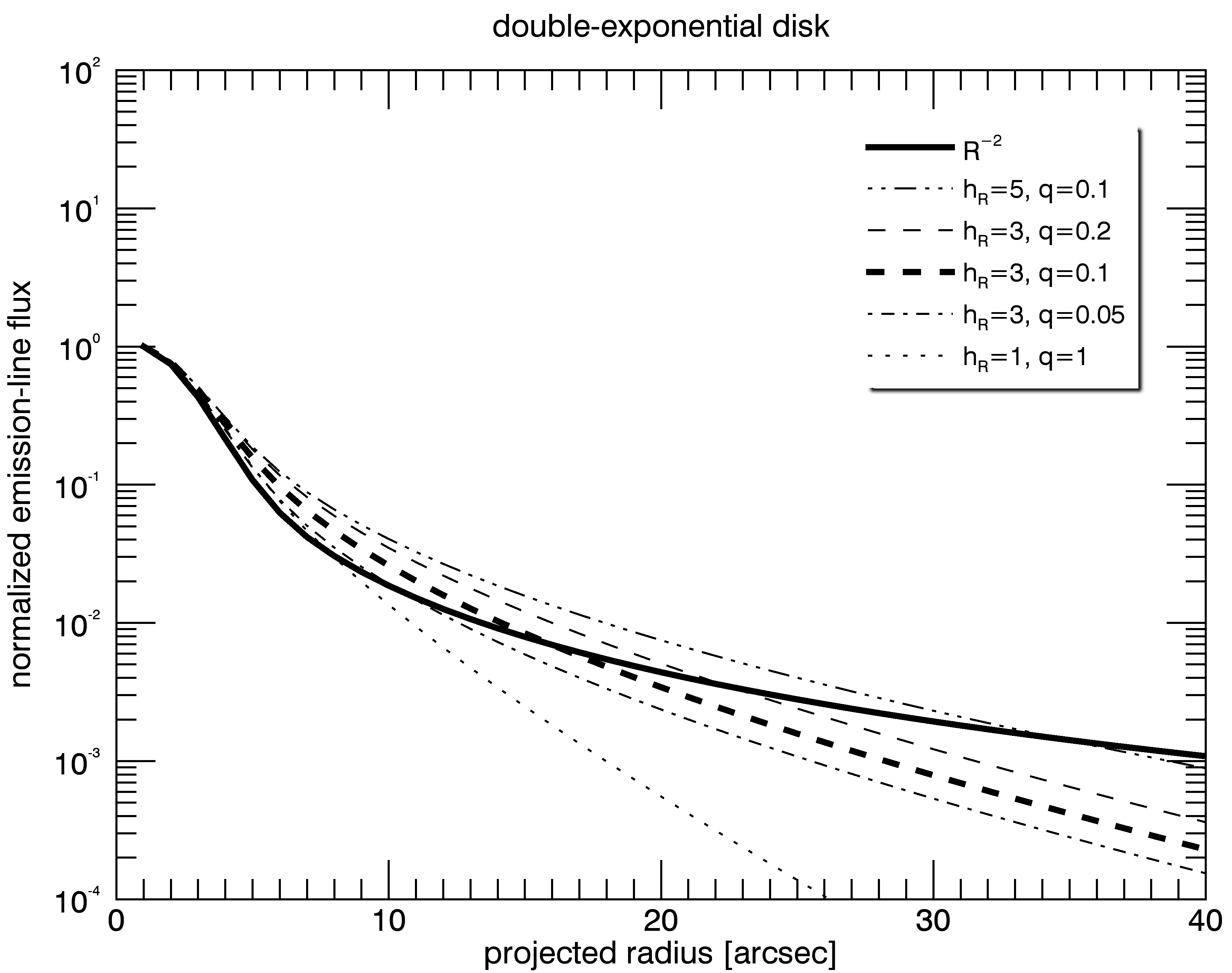}
\end{center}
\caption{
Normalised flux profiles as a function of projected radius $R$ of gas being ionised by a central point source. The thick solid curve assumes that the gas is optically thin and distributed in an infinitesimally thin disk with a constant filling factor and constant density, so that the fall-off is the same inversely squared with radius as the point-source radiation. The thick dashed curve is when the gas is distributed in an axisymmetric disk of finite thickness with gas density both radially and vertically declining exponentially as $\propto \exp[-R/h_R] \exp[-|z|/(q \, h_R)]$ with fiducial values for the scale length of $h_R = 3$\,kpc and for the flattening of $q=0.1$, viewed at an inclination angle of $i=60^\circ$. The thin long/short dashed curves show the effect of a factor of two thicker/thinner disk, the dash-dotted curve is for when the scale length is much larger, and the dotted curve is when the gas distribution is spherical and more centrally concentrated.
}
	\label{fig:pointsource}
\end{figure}

We illustrate the latter by a simple model in which a central point source
ionises optically thin gas with a constant filling factor distributed in an
axisymmetric disk viewed at an inclination angle $i=60^\circ$ (with
$i=0^\circ$ face-on and $i=90^\circ$ edge-on). Combined neutral and molecular
hydrogen measurements in nearby galaxies show that the gas density declines
exponentially in radius \citep{bigiel2012} and that the vertical fall-off is
typically well matched by an exponential as well. Henceforth, we adopt a
double-exponential for the gas density $\propto \exp[-R/h_R] \exp[-|z|/(q \,
h_R)]$ with fiducial values for the scale length of $h_R = 3$\,kpc and for
the flattening of $q=0.1$.
The resulting flux profile is shown in Fig.~\ref{fig:pointsource} as the
thick dashed curve, whereas the thin long/short dashed curves show the effect
of a factor two thicker/thinner disk. The differences with respect to the
$1/R^2$ fall-off (thick solid curve) are much smaller than the offset from the
on average, much shallower observed flux profiles shown in Fig.~\ref{fig:fig3}.
The same holds true for a thin disk ($q=0.1$) with a much larger scale length
($h_R = 5$\,kpc), as indicated by the dash-dotted curve, or a spherical ($q=1$) and
more centrally concentrated ($h_R = 1$\,kpc) gas distribution, represented by the dotted curve.

\subsection{Impact of geometric projections}

Even if the disks of galaxies are intrinsically round, the inclination at which we observe them results in projection effects that act both on the minor axis radius coordinate and the effective gas density and hence, line emitting region.
Under the assumption of a geometrically thin gas distribution, a
proper de-projected radius would be described by $\sqrt{R_a^2+(R_b/\cos i)^2}$
with a major-axis radius component $R_a$, observed minor axis component $R_b$,
and inclination angle $i$.
The gas distribution itself, on the other hand, would be projected by the same
amount as the minor axis component, $\cos i$. 
Hence, the observed projected
profiles have all data points moved to smaller observed radii by different
amounts, while the flux density is moved to higher values. 

\begin{figure}
	\includegraphics[width=\columnwidth]{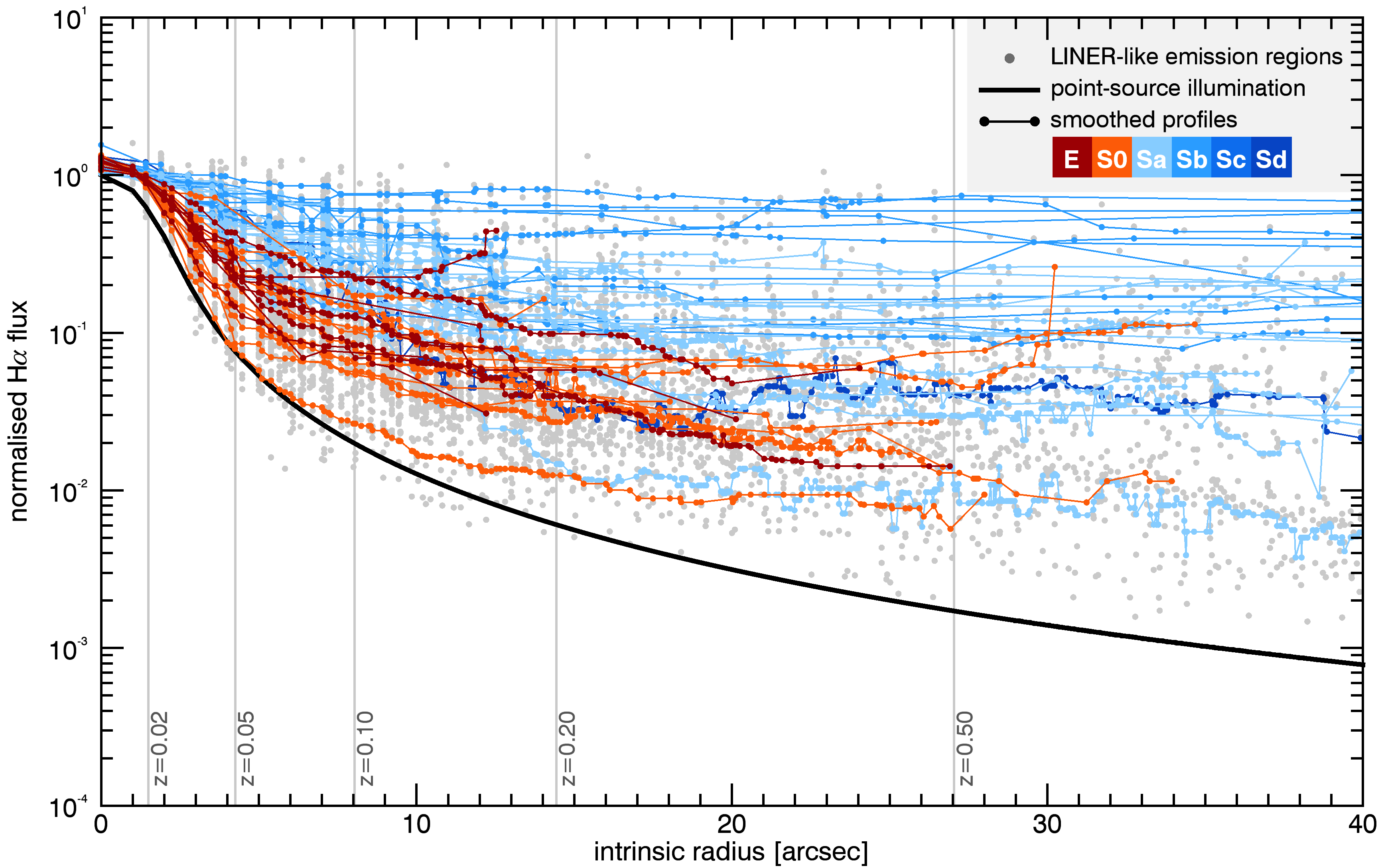}
	\caption{De-projection of radial H$\alpha$ surface brightness
          profiles. To demonstrate how strongly potential projection effects
          of galaxy inclination could impact our results, all disk-dominated
          galaxies of the sample were subjected to a maximal geometrical
          de-projection. For this test, it was assumed that these galaxies were
          infinitely thin disks and that observed ellipticities were fully due
          to an inclination of the disk with respect to the observer's line of
          sight. Both radius and gas surface density projection were considered.
          The comparison to Fig.~\ref{fig:fig3} shows a
          stretch of the radius axis for some objects, but neither qualitative
          nor quantitative difference in surface brightness excess for the
          galaxies over the point-source line is shown.}
	\label{fig:deprojected}
\end{figure}

To assess whether this produces a significant net increase or decrease of the difference
between observations and models as seen in Fig.~\ref{fig:fig3}, a
tentative and maximal de-projection of all disk-dominated galaxies in our sample was
carried out for illustrative purposes, as shown in
Fig.~\ref{fig:deprojected}. 
For this calculation, we adopted $\cos i = 1 - \epsilon$ for a thin disk with observed ellipticity $\epsilon$
that is derived from an isophotal analysis of the SDSS images.
As can be seen, the impact on the discrepancy model--observations is at most
small, and it can be concluded that projection effects play no significant
role in the interpretation.

\subsection{ Impact of mixed-in star formation contribution}

A selection of spaxels in a BPT diagram above the theoretical upper limit of
where star formation alone can produce given line ratios does not preclude a
significant contribution to the line emission from star formation (SF). In
principle, a mix of a fiducial ``pure LINER'' with a ``pure SF'' ionising
radiation field, can lead to substantial SF contribution to the emission line
flux outside the classical SF region. Given that early-type galaxies with
negligible SF show similarly shallow flux profiles as late-type galaxies with
significant SF (Fig.~\ref{fig:fig3}), this result already hints that mixed-in SF
contributions cannot be the source of the discrepancy with a point-source
illumination.

\begin{figure}
\begin{center}
	\includegraphics[width=\columnwidth]{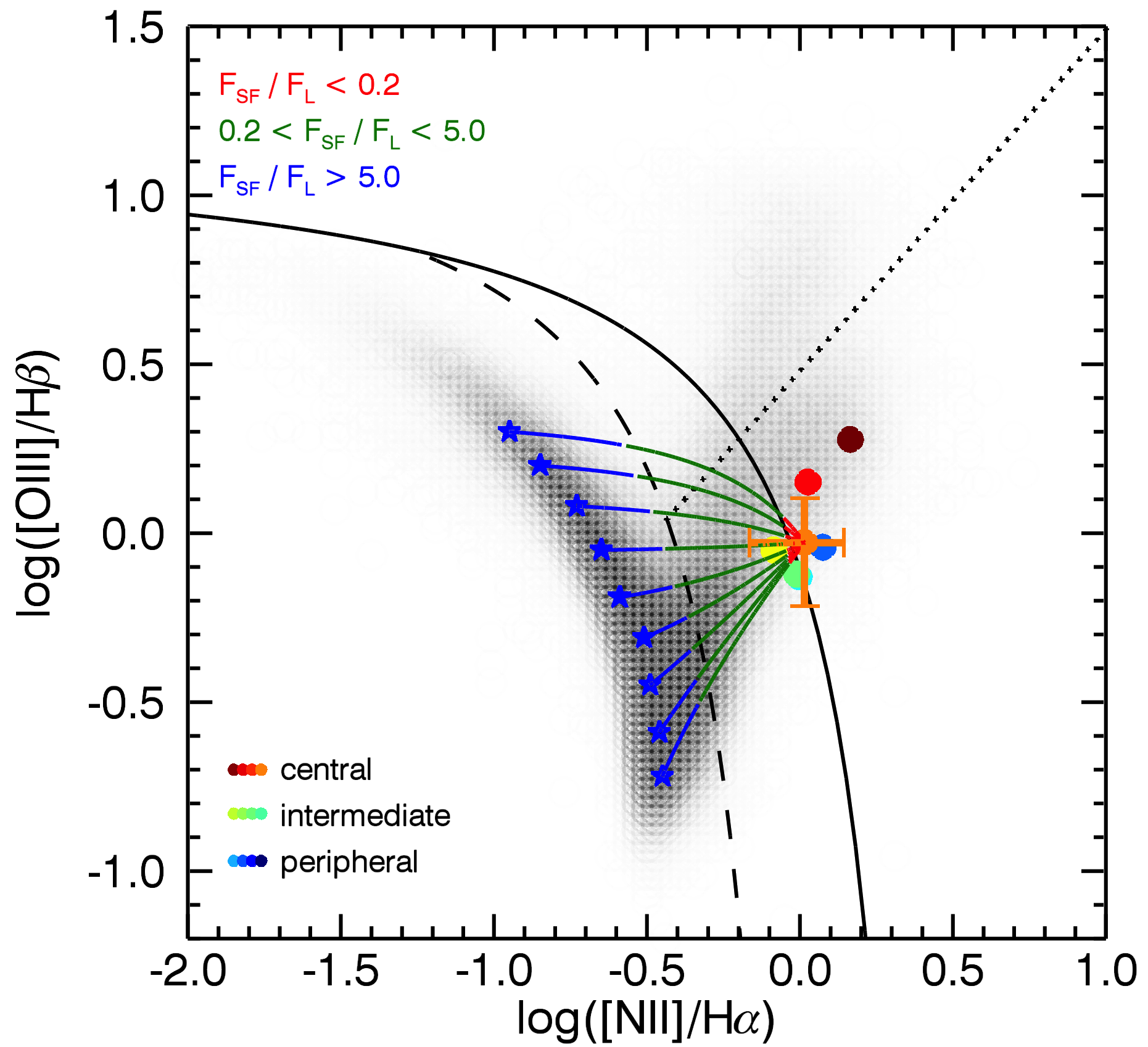}
\end{center}
\caption{
Resulting line ratios if line emission flux purely from SF (blue stars) and
purely from a LINER mechanism (orange point) are linearly superposed (coloured
lines). The generating H$\alpha$ flux ratios are colour coded. This shows that the SF
contribution in the selected LINER regime right of the solid curve is negligible ($F_{SF}$/$F_L$ < 0.2)
and cannot be generating the discrepancy seen in radial emission line profiles with respect to a
point-source illumination of $>$1\,dex at larger radii.
}
	\label{fig:SFcontribution}
\end{figure}

Even so, we estimate the level of mixed-in SF contribution by assuming that
the observed emission-line fluxes are a linear combination of flux $F_L$ from
``pure LINER'' ionisation and flux $F_{SF}$ from ``pure SF'' ionisation. The line
ratios from these pure ionisation sources correspond to points in the LINER
and SF regimes of the BPT diagrams. For a given H$\beta$-to-H$\alpha$ flux ratio, the so-called Balmer decrement, which increases
the ratio $F_{SF}/F_L$ from zero, traces a curve from the pure LINER
point toward the pure SF point in the BPT diagram, as illustrated in
Fig.~\ref{fig:SFcontribution}.

As the latter pure SF points (blue stars), we use
locations on the SF-ridge of SDSS galaxies \citep{kewley2006} that are shown as grey
levels in the background. We infer to the pure LINER point (orange point), 
from the average position of the nine elliptical LINER galaxies from our
sample. These elliptical galaxies are devoid of any SF, but there could still be
mixed-in ionisation contribution from a central AGN in their inner regions.
Indeed, computing the position based on emission-line fluxes from
different galactocentric annuli shows that the two central-most annuli
(dark-red and red points) yield a position more toward the Seyfert regime. The
average position resulting from the annuli further out nicely converge to the
same position, again in-line with ionisation by the same but non-central sources.

To go from this pure LINER point to the black solid demarcation line, the
SF-to-LINER flux ratio increases to $F_{SF}/F_L=0.2$, so that a maximum of
1/6th of observed flux could be due to mixed-in SF contribution. The resulting
decrease in the observed flux is negligible with respect to the offset from the
point-source illumination in Fig.~\ref{fig:fig3}. Placing the pure
LINER point further away from the demarcation line increases the
possible SF-to-LINER flux ratio. Given the spread in the convergence point
among the elliptical galaxies (orange cross), however, we find
$F_{SF}/F_L<0.5$. Hence only 1/3rd of the observed flux could still be
due to mixed-in SF contribution, and we conclude that, SF cannot be a significant
source of the observed shallow emission-line flux profiles even for the spiral galaxies,
whereas ionisation from the same common old stars forms the natural explanation.

\section{Discussion and Conclusions}
\label{sec:discussion}

For 48 galaxies with LINER-like emission we unambiguously show, that the class of LINER galaxies, contrary to their 30-year old paradigm, are not predominantly powered by a central AGN, since their radial emission-line surface brightness profiles are inconsistent with ionisation by a central point-source and hence cannot be due to an AGN alone.
When using this result in reverse, we conclude that the power source for
LINER-like emission must be extended, which is possibly distributed all through the
galaxy, while it is clear that an adequate supply of gas is indispensable for such line emission to exist in the first place.
Despite its name LINER-like emission covers all regions of the
galaxies, aside from those where star-formation dominates the emission.
There is also no spatial collimation as would be expected in the case of shock-driven
ionisation.  Henceforth, the most probable energy source are hot evolved stars
after their asymptotic giant branch (AGB) phase. This was already suggested
before \citep{binette1994, goudfrooij1997, goudfrooij1999},
but models of this phase in stellar evolution have
only recently been picked up again \citep{stasinska2008}. After stars leave
their main sequence of hydrogen burning, and after a few subsequent evolutionary stages,
they enter the so-called AGB. In the following post-AGB phase, the stars can
become sufficiently hot to produce a spectrum capable of ionising atoms with a
substantial ionisation potential. Realising this has the perplexing implication that
every galaxy for both early- and late-type galaxies with stellar populations
older than $\sim$1 Gyr must have a radiation field from post-AGB stars that
can ionise at least part of the interstellar gas when present.

  Even if central or extended LINER-like emission is predominantly powered by
  post-AGB stars, it does not preclude the existence of AGNs in LINER galaxies.
  The AGN could provide some of the central radiation in some LINER
  galaxies, which may even host a higher fraction of AGN compared
  to the general population of massive galaxies \citep{gonzalez2009}: The
  reason for this could in the simplest case be a selection effect due to the
  required presence of central reservoirs of gas and hence potential for it
  being accreted onto a central black hole. However, many galaxies with
  LINER-like emission do not have a central AGN. This shows that the LINER
  diagnostic is in general not a good predictor for the presence of an AGN.
  
  In addition, relying on LINER signatures for AGN selection suffers from
  aperture effects: Observing galaxies at different distances therefore different
  physical apertures (see Fig.~\ref{fig:fig3}) clearly makes a comparison of
  the properties even of classical LINERs difficult because of the mixing of
  signals from emitting regions at different radii.

  This emphasises that LINERs or galaxies with widespread LINER-like emission are not
  just a mixed bag of properties but are mainly just normal galaxies with some
  gas content in the absence of substantial ionisation fields from young
  stars and AGN.

The consequences are profound for different fields in astrophysics, ranging
from galaxy evolution models, which have hot old stars as an always-present
ionisation source that create LINER-like emission whenever gas is present, to
black hole studies, which do not have to resort to rare accretion models to
explain the LINER galaxies.  The three immediate consequences from this result
are described below.

First, the ubiquitous presence of ionising radiation from post-AGB stars means that
\emph{galaxies with LINER-like emission are not a
class defined by a property, but rather by the absence of a property}, or
the absence of a stronger radiation field, as for example produced by young stars. This both
explains why typical LINER galaxies are massive and old: These
are the only galaxies without substantial star formation and with
enough post-AGB stars to generally detect the LINER signature \citep{papaderos2013}.
This also tells us why LINERs appeared as a mixed bag: The presence of other sources of
energy -- AGN, star formation, shocks  -- is actually completely
independent of the source powering the LINER signature.

Second, we need to revisit the properties of classical AGN host galaxies,
since a number of studies in the past decade used, for example, the SDSS survey to
investigate AGN host galaxies in the local Universe. Since the sample
selection LINERs outnumber classical Seyferts 5:1, were often counted
into the AGN class \citep{kauffmann2003,kauffmann2009,schawinski2007}, and
were being impacted by the different physical apertures
covered at different redshifts (Fig.~\ref{fig:fig3}), these studies might have
come to biased results.

Third, with LINER-signatures now being removed as a self-contained AGN indicator,
the family of AGN becomes much smaller and simpler.

\begin{acknowledgements}
  The authors would like to thank all of the CALIFA collaboration for their
  input, Brent Groves for very useful discussions on ionisation
  properties and Remco van den Bosch for his technical and scientific advises.
  RS acknowledges support by the IMPRS for Astronomy \& Cosmic
  Physics at the University of Heidelberg.
  KJ is supported by the Emmy Noether-Programme of the German Science
  Foundation DFG under grant Ja~1114/3-2 and the German Space Agency DLR.
  G.~v.~d.~V.\ and J.~F.-B.\ acknowledge the DAGAL network from the People 
  Programme (Marie Curie Actions) of the European Union's Seventh Framework
  Programme FP7/2007-2013/ under REA grant agreement number PITN-GA-2011-289313.
  J.~F.-B. further acknowledges financial support from the Ram\'on y Cajal Program and 
  grant AYA2010-21322-C03-02 from the Spanish Ministry of Economy and Competitiveness (MINECO).
  VW acknowledges support from the ERC Starting Grant SEDmorph.
  R.~A. Marino was also funded by the spanish programme of International Campus of Excellence Moncloa (CEI).
This study makes uses of the data provided by the Calar Alto Legacy
  Integral Field Area (CALIFA) survey (http://califa.caha.es/) and is based on
  observations collected at the Centro Astron\'omico Hispano Alem\'an (CAHA)
  at Calar Alto, operated jointly by the Max-Planck-Institut f\"ur Astronomie
  and the Instituto de Astrof\'isica de Andaluc\'ia (CSIC).

\end{acknowledgements}

\bibliographystyle{aa}
\bibliography{liner}
\end{document}